\documentclass[preprint,prd,nofootinbib,showpacs,rotate]{revtex4}

\usepackage{graphicx}
\usepackage{dcolumn}
\usepackage{bm}
\input{epsf}
\begin{document}

\title{Glueball Regge trajectories from gauge/string duality and the Pomeron}

\author{Henrique Boschi-Filho}
\email{boschi@if.ufrj.br}
\affiliation{Instituto de F\'{\i}sica, 
Universidade Federal do Rio de Janeiro, Caixa Postal 68528, RJ 21941-972 -- Brazil}
\author{Nelson R. F. Braga}
\email{braga@if.ufrj.br}
\affiliation{Instituto de F\'{\i}sica,
Universidade Federal do Rio de Janeiro, Caixa Postal 68528, RJ 21941-972 -- Brazil}
\author{Hector L. Carrion}
\altaffiliation{ Present address: Instituto de F\'{\i}sica, Universidade de S\~ao Paulo, 
Caixa Postal 66318, 05315-970, S\~ao Paulo-S.P., Brasil}
\email{mlm@if.ufrj.br}
\affiliation{Instituto de F\'{\i}sica,
Universidade Federal do Rio de Janeiro, Caixa Postal 68528, RJ 21941-972 -- Brazil}

\begin{abstract} 
The spectrum of light baryons and mesons has been reproduced recently by 
Brodsky and Teramond from a holographic dual to QCD inspired in 
the AdS/CFT correspondence.
They associate fluctuations about the AdS geometry 
with four dimensional angular momenta of the dual QCD states. 
We use a similar approach to estimate masses of glueball states with different spins 
and their excitations. We consider Dirichlet and Neumann boundary conditions
and find approximate linear Regge trajectories for these glueballs.
In particular the Neumann case is consistent with the Pomeron trajectory.

\end{abstract}

\pacs{ 11.15.Tk ; 11.25.Tq ; 12.38.Aw ; 12.39.Mk .}

\maketitle

\section{Introduction}
The observation that hadrons show up in  approximate linear Regge
trajectories was one of the initial motivations for developing string theory.
Recently very good estimates for masses of light mesons and baryons
were obtained from string theory in a sliced $AdS_5 \times S^5\,$
spacetime\cite{deTeramond:2005su}.
On the other side, experimental results for the cross sections of soft processes
show an increase with energy corresponding to pomerons 
with Regge trajectories of the form\cite{Landshoff:2001pp}
\begin{equation}
\label{Pomeron}
\alpha (t = M^2\,) \,\approx \,  1.08 \,+\, 0.25\, M^2\,
\end{equation}
\noindent where masses are in GeV. Furthermore, 
it has been suggested that pomerons may be related to glueballs. 

We use a similar approach to that of ref.\cite{deTeramond:2005su} to 
estimate masses of glueballs 
with different angular momenta and obtain the corresponding Regge trajectories.
Our results for the glueball trajectories show consistency with that 
of soft Pomerons. 

Strong interactions are well described by QCD (Yang Mills SU(3) plus
fermionic matter fields). In the high energy regime one can perform perturbative 
calculations.  At low energies QCD is non perturbative 
and the usual approach is to consider QCD in a lattice.  
In particular a lattice analysis of the consistency of glueball Regge trajectories
with pomerons has been done recently\cite{Meyer:2004jc}. 

An alternative approach is to consider a dual description
of strong interactions in terms of string theory.
A connection between SU(N) gauge theories (for large N) and string theory
was realized long ago by 't Hooft\cite{'tHooft:1973jz}.
A very important recent result relating gauge and string theories
was obtained by Maldacena\cite{Malda}. He established a 
correspondence between string theory in $AdS_5 \times S^5 \,$ space-time
and ${\cal N } = 4$ Super Yang Mills SU(N) theory for large N 
in its four dimensional boundary. This super Yang Mills theory is conformal. 
Soon after, a proposal of a correspondence closer to QCD 
(non conformal and non supersymmetric) was discussed by Witten\cite{Wi2}. 
In this formulation QCD would be described by string theory in an AdS-Schwazschild 
black hole metric.
Glueball masses were estimated in this context, using a WKB approximation in 
\cite{MASSG,MASSG2,MASSG3,MASSG4,Constable:1999gb,MASSG5,MASSG6,MASSG7}.
Also there are many interesting estimates of glueball masses from 
dualities involving different geometries generated by string theory
as for example\cite{Caceres:2000qe,ACEP,Amador:2004pz,Evans:2005ip,
Caceres:2005yx}.

A phenomenological approach to estimate hadron masses inspired in 
the AdS/CFT correspondence was proposed in \cite{Boschi-Filho:2002vd,Boschi-Filho:2002ta}
applied to the case of scalar glueballs.
An energy scale was introduced in analogy with the discussion
of hard scattering from AdS/CFT in \cite{PS} (see also\cite{Polchinski:2002jw}).  
In this approach  supergravity fields in an $AdS_5 \,$ slice times a compact
$S^5$ space 
are considered
as an approximation for a string theory  dual to QCD. 
The metric of this space can be written as
\begin{equation}
\label{metric3}
ds^2=\frac { R^2 }{ z^2}\Big( dz^2 \,+(d\vec x)^2\,- dt^2 \,\Big)\,+ 
R^2 d \Omega_5^2 .
\end{equation}

\noindent where the size of the slice: $0 \le z \le z_{max}\,$
is related to the QCD scale
\begin{equation}
\label{zmax}
z_{max}\,=\,{1\over \Lambda_{QCD}}\,.
\end{equation}
 
In this phenomenological approach Dirichlet boundary conditions 
were imposed at $ z = z_{max}$ and
the ratios of the masses of the scalar glueball
$0^{++}\,$ and its spinless excitations were 
obtained\cite{Boschi-Filho:2002vd,Boschi-Filho:2002ta}.
These results are in good agreement with lattice and AdS-Schwazschild 
results. Note that we do not consider excitations in the $S^5$ directions 
since according to the AdS/CFT correspondence they are related to the supersymmetric
structure of the boundary theory.
For other results related
to strong interactions  from AdS/CFT see also \cite{GI,BB3,BT,AN,Brodsky:2003px,
PandoZayas:2003yb,
AN2,AN3,Andreev:2004sy,Bigazzi:2004ze,Erlich:2005qh,DaRold:2005zs}. 

\section{Glueball Masses }

In ref. \cite{deTeramond:2005su} very interesting results for the hadronic
spectrum were obtained considering scalar, vector and fermionic fields 
in the sliced $ AdS_5 \times S^5 $ space.
It was proposed that massive  bulk states corresponding to fluctuations about 
the $AdS_5$ metric are dual to QCD states with  
angular momenta (spin) on the four dimensional boundary.

According to the AdS/CFT correspondence, massless scalar string states 
are dual to boundary scalar glueball operators\cite{GKP,Witten:1998qj}.
On the other hand, scalar string excitations with mass $\mu$ 
couple to boundary operators with dimension $\Delta =  2 + \sqrt{ 4 + (\mu R )^2 }\,\,$.
This happens because these massive states behave as $ z^{4 - \Delta}\,$ 
near the AdS boundary (small $ z $).  
Scalar glueball operators ${\cal O}_4\,=\, F^2$ have dimension 4, while 
glueballs operators ${\cal O}_{4 + \ell}\,=\, F D_{\{\mu_1}...D_{\mu_\ell\,\}} F\,$ 
with spin $\ell\,$ have dimension $ 4 + \ell \,$.
Then a consistent coupling between string states with mass $\mu$ and glueball operators
with spin $\ell$ requires that  
\begin{equation}
\label{MASSA}
( \mu R )^2 \,=\, \ell (\ell + 4 ) .
\end{equation}

\noindent This means that the masses of these AdS modes have a discrete spectrum
since they are in correspondence with glueball operators of integer spin.

We will assume that such dualities established in the AdS/CFT correspondence 
are approximately valid in our phenomenological model of an AdS slice. 
So we take glueball operator with spin $\ell$ to be dual to massive string 
states, with mass given by eq. (\ref{MASSA}), in the AdS slice.

The glueball operators in the AdS/CFT correspondence are all massless
respecting conformal invariance. Once we introduce a size $z_{max}\,$ in the 
AdS space there will be an infrared cut off in the boundary, which we identify with
$ \Lambda_{QCD}$, breaking conformal invariance. 
The presence of the slice implies an infinite tower of discrete modes  
in the $z$ direction for the bulk states. This discretization does not alter 
the asymptotic behavior (small $z$) of bulk modes which is related 
to their mass $\mu$. We assume that these bulk discrete modes in the $z$ direction 
are related to the masses of the non conformal glueball operators.
Using this model we will calculate glueball masses and the corresponding 
Regge trajectories.
 
The solutions for scalar fields with mass $\mu$ in $AdS_5$  
satisfy\cite{GKP,Witten:1998qj}
\begin{equation}
\Big[ z^3 \partial_z {1\over z^3} \partial_z \,+\, \eta^{\mu\nu} \partial_\mu 
\partial_\nu \,-\, { (\mu R )^2  \over z^2 } \,\,\Big] \,\phi \,\,=\,\,0\,\,.
\end{equation}

\noindent
Considering plane wave solutions in the four dimensional coordinates 
$\vec x $ and $t$ for states with mass $\mu$ given by  eq.  (\ref{MASSA}) 
one can write the solutions as
\begin{equation}
\phi\,( x, z )\,=\,C_{\nu , k}\,e^{-i P.x} \,z^2 J_{\nu} ( u_{\nu , k} \, \,z\,)\,, 
\end{equation}

\noindent where $\nu = 2 \,+\, \ell \,$ and the discrete modes $ u_{\nu , k}\,$ 
($ k =1,2,... )\,$ are determined by the boundary conditions. 
Here we will consider two possibilities: 
\begin{eqnarray}
\phi (\,z =z_{max}\,) &=& 0 \qquad \qquad ({\rm Dirichlet})\,; \\
\partial_ z \phi \vert_{z =z_{max}} &=& 0 \qquad  \qquad ({\rm Neumann })\,\,.
\end{eqnarray}

\subsection{Dirichlet boundary conditions}

In the case of Dirichlet boundary conditions, as used in
references \cite{deTeramond:2005su} and 
\cite{Boschi-Filho:2002vd,Boschi-Filho:2002ta} one obtains 
\begin{equation}
\label{Dmasses}
u_{\nu , k} \,=\, {\chi_{_{\nu , k}} \over z_{max}}\,=\,\chi_{_{\nu , k}}\,\Lambda_{QCD}
\qquad ;\qquad J_\nu (\,\chi_{_{\nu , k}}\,) \,=\,0.  
\end{equation}
  
Assuming the duality between these modes in the  
$AdS_5 $ slice and the glueball operators, 
the scalar glueball $0^{++}$ is related to the massless scalar. 
So its mass is proportional to $\chi_{_{2 , 1}}$. 
The excited scalar glueball states
$0^{++\ast ... \ast}$ correspond to the other values of $k$ and their masses are 
proportional to  $\chi_{_{2 , k}}$.

The higher angular momenta glueballs $J^{++}$ are related to the massive scalar modes
according to $ ( \mu R )^2 \,=\, \ell (\ell + 4 )\,$ with $ \ell = J$.
Then the mass of the lightest state with angular momentum $J$ is proportional to  
$\chi_{_{\,2+\ell \, ,\, 1}}\,$.
The corresponding excitations are proportional to $\chi_{_{\,2+\ell \, ,\, k}}\,$.

We show in Table I the results for even angular momenta that may be related 
to the phenomenological pomeron which has a trajectory with even signature.
We introduced the mass of the lightest glueball as an input and found  
the glueball spectrum from equation (\ref{Dmasses}). 
This input for the lightest glueball is in accordance with lattice 
results\cite{LAT1,LAT2}.
The results for the excitations of $0^{++}\,$ were obtained previously in 
\cite{Boschi-Filho:2002vd,Boschi-Filho:2002ta} and are in good agreement with
the masses estimated using AdS-Schwazschild 
black hole metric\cite{MASSG,MASSG2,MASSG3,MASSG4,MASSG5,MASSG6,MASSG7}.

Our result for the ratio of masses $M_{2^{++}} /M_{0^{++}}\,=\,1.48\,\,$  is in good
agreement with lattice \cite{LAT1,LAT2} and deformed conifold 
results \cite{Amador:2004pz}.

\begin{table}
\centering
\begin{tabular}{ | c | c | c | c |} 
\hline 
Dirichlet $\,\,$ 
 & $\,\,$ lightest $\,\,$   &
$1^{st}$ excited $\,\,$ & $2^{nd}$ excited \\
glueballs $\,\,$     &   state & state & state \\
 \hline
 $0^{++}$ & 1.63  &  2.67 & 3.69 \\ 
 $2^{++}$ & 2.41    & 3.51  & 4.56  \\
 $4^{++}$ & 3.15  & 4.31  & 5.40  \\
 $6^{++}$ & 3.88  & 5.85  & 6.21 \\
 $8^{++}$ & 4.59 & 5.85  & 7.00 \\
 $10^{++}$ & 5.30 & 6.60   & 7.77 \\
\hline
\end{tabular}
\parbox{5in}{\caption{ 
\sl Masses of glueball states $J^{PC}\,$ with even $J$ expressed in GeV,
estimated using the sliced $AdS_5\times S^5$ space with Dirichlet boundary conditions.
The mass of $0^{++}$ is an input from lattice results \cite{LAT1,LAT2}.
}}
\end{table}

\bigskip

\subsection{Neumann boundary conditions}

\bigskip 
Considering Neumann boundary conditions, the vanishing of the scalar field derivative 
at $z_{max}\,$ leads to 
\begin{equation}
( 2 - \nu )  J_\nu (\,\xi_{\,\nu , k}\,) \,
+\,\xi_{\,\nu , k}\,J_{\nu -  1}  (\,\xi_{\,\nu , k}\,) 
\,=\,0.  
\end{equation}

\noindent The correspondence between QCD and scalar string states is taken exactly as 
in the Dirichlet case. The glueball masses are now given by
\begin{equation}
u_{\nu , k} \,=\, {\xi_{_{\nu , k}} \over z_{max}}\,=\,\xi_{_{\nu , k}}\,\Lambda_{QCD}
\end{equation}
  
In this case we also take the mass of the lightest glueball as an input. 
The results for states with even spin are shown in Table II. 

Here in the Neumann case the ratios of the masses 
\begin{eqnarray}
{ M_{2^{++}} \over M_{0^{++}}} &=& 1.56 \\
{ M_{0^{++\ast}} \over M_{0^{++}}} &=& 1.83 
\end{eqnarray}

\noindent are in very good agreement with lattice\cite{LAT1,LAT2} and deformed conifold 
results\cite{Amador:2004pz}.

\begin{table}
\centering
\begin{tabular}{ | c | c | c | c |} 
\hline 
Neumann $\,\,$ 
 & $\,\,$ lightest $\,\,$   &
$1^{st}$ excited $\,\,$ & $2^{nd}$ excited \\
glueballs $\,\,$     &   state & state & state \\
 \hline
 $0^{++}$ & 1.63  & 2.98  & 4.33  \\ 
 $2^{++}$ & 2.54    & 4.06 & 5.47  \\
 $4^{++}$ & 3.45 & 5.09  & 6.56 \\
 $6^{++}$ & 4.34  & 6.09 & 7.62 \\
 $8^{++}$ & 5.23 & 7.08 & 8.66 \\
 $10^{++}$ & 6.12 & 8.05  & 9.68 \\
\hline
\end{tabular}
\parbox{5in}{\caption{ \sl Masses of glueball states $J^{PC}\,$ with even $J$ 
expressed in GeV, estimated using the sliced $AdS_5\times S^5$ space with 
Neumann boundary conditions.
The mass of $0^{++}$ is an input from lattice results \cite{LAT1,LAT2}.
 }}
\end{table}

\section{Regge trajectories}

From the results of tables I and II one finds relations between spin and mass  squared
for the glueballs that represent the corresponding Regge trajectories.
We find that these trajectories for the glueballs are non-linear.  
This is in agreement with general properties of Regge trajectories, as discussed 
for example in \cite{Tang:2000tb}.

In order to compare these results with the Pomeron behavior of 
eq. (\ref{Pomeron}) it is interesting to consider linear 
approximations for these trajectories as in ref. \cite{Landshoff:2001pp}

\begin{equation}
 J \,=\,\alpha (t = M^2\,)\,=\, \alpha_0 \,+\,\alpha^{\prime} \, M^2\,.
\end{equation}

\noindent Here we will be interested in the trajectories of the glueball
lightest states with even $J$ only. Also, as discussed in \cite{Meyer:2004jc} 
the $0^{++}\,$ glueball 
is not expected to contribute to the Pomeron trajectory that has a positive intercept.
For more discussions on the Pomeron intercept see for instance \cite{Luna:2004gr}. 
So we will consider linear fits for the states $2^{++}, 4^{++} ,...\,$ for both 
Dirichlet and Neumann boundary conditions. 

In particular, for the Neumann case the results are compatible with the Pomeron  
trajectory. For instance, for the states  
$J^{++}\,$ with $J = 2,4,...,10\,$ we find 
\begin{equation}
\alpha^{\prime}\,=\,(\,0.26 \pm 0.02 \,)GeV^{-2} 
\qquad ; \qquad \alpha_0 \,=\,0.80 \pm 0.40
\end{equation}

\noindent This trajectory is shown in figure {\bf 1}. Note that we 
are not considering errors in the masses of the Glueballs. The errors appearing
in the estimated coefficients $\alpha^{\prime}\,$ and $ \alpha_0 \,$ refer to 
the deviations with respect to the linear fit.

For other set of points we also find results compatible with the Pomeron trajectory.
In particular for the set of states $4^{++}\,,6^{++}\,,8^{++}\,$ we find
$ \alpha^{\prime}\,=\,(0.26\, \pm 0.01  \,)GeV^{-2} \,$ and
$\, \alpha_0 \,=\,1.01 \pm 0.30\, $ .

We note that the slope $\alpha^{\prime}$ in the Neumann case does not
vary considerably with the set of states considered in the linear
approximation. The error of the slope is still small 
and consistent with the Pomeron result.
 
For the Dirichlet case, taking the states 
$J^{++}\,$ with $J = 2,4,...,10\,$ 
we find a linear fit with 
\begin{equation}
\alpha^{\prime}\,=\,( \, 0.36 \pm  \,0.02\,)\,GeV^{-2}\, \qquad ; \qquad
 \alpha_0 \,=\,0.32 \pm 0.36 \,.
\end{equation}

\noindent  These states and the corresponding linear fit
are shown in figure {\bf 2}. 
The slope of this trajectory is higher than the Pomeron result in eq.(\ref{Pomeron}).
For other sets of states using Dirichlet boundary condition 
we also find linear approximations with slopes which are higher than that of the Pomeron.
For instance, with $J = 4,...,10\,$ we find
$ \alpha^{\prime}\,=\,(0.33\, \pm 0.02  \,)GeV^{-2} \,$ and
$\, \alpha_0 \,=\,0.90 \pm 0.32$. 

\begin{figure}
\centering
\includegraphics[width=8cm]{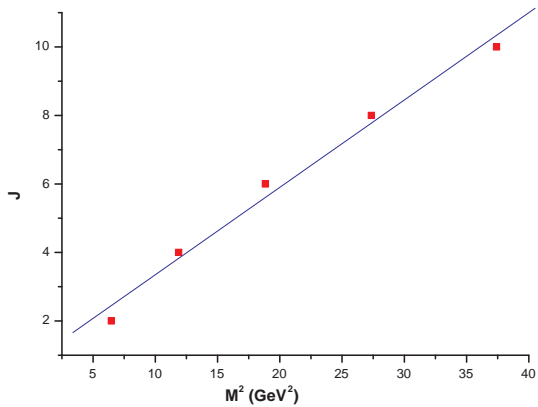}
\parbox{5in}{\caption{Approximate linear Regge trajectory for Neumann Boundary 
condition for the states
$\,\,2^{++}\,,4^{++}\,,6^{++}\,,8^{++}\,,10^{++}\,$.}}
\end{figure}

\begin{figure}
\centering
\includegraphics[width=8cm]{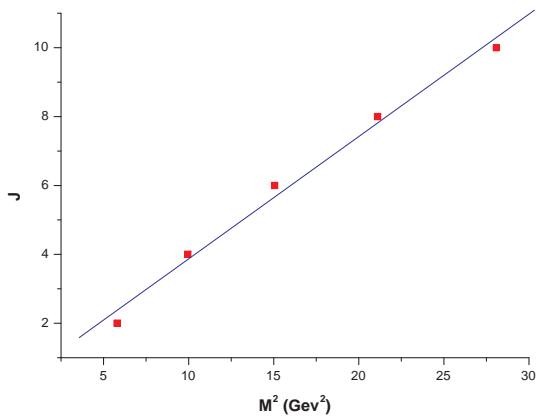}
\parbox{5in}{\caption{Approximate linear Regge trajectory for Dirichlet 
Boundary condition 
for the states $\,\,2^{++}\,,4^{++}\,,6^{++}\,,8^{++}\,,10^{++}\,$.}}
\end{figure}

\section{Conclusion}
We found simple estimates for masses of glueballs of different spins 
in a sliced $AdS_5 \times S^5\,$ inspired in the AdS/CFT duality. 
These results are in good agreement with other estimates in the literature.
It is remarkable that for the case of Neumann boundary condition the linear 
approximation for glueball Regge trajectories
\begin{equation}
\alpha (t = M^2) \,=\, (\,0.80 \pm 0.40\,)
\,+\, (\,0.26 \pm 0.02 \,)M^{2} 
\end{equation}

\noindent  is consistent with the 
Pomeron trajectory of eq.(\ref{Pomeron}). 

This result shows that the Neumann boundary condition 
seems to work better than Dirichlet for glueballs in this holographic model.
Both choices correspond to vanishing flux for bulk scalar fields 
at $z = z_{max}\,$ and  would be physically acceptable conditions.
It is interesting to note that similar Neumann conditions appear in the Randall Sundrum
model \cite{Randall:1999ee} as a consequence of the orbifold condition.
   
\bigskip

\noindent {\bf Acknowledgments}: 
We would like to thank Guy de Teramond and Stanley Brodsky for important discussions.
The authors are partially supported by CNPq, CLAF and Fapesp. 
H. L. C. would like to thank CBPF for hospitality during part of this work.


\begin{thebibliography}{30}

\bibitem{deTeramond:2005su}   G.~F.~de Teramond and S.~J.~Brodsky,
Phys. Rev. Lett. {\bf 94} (2005) 201601.

\bibitem{Landshoff:2001pp}
  P.~V.~Landshoff,
  ``Pomerons,'', published in  ``Elastic and Difractive Scattering" 
 Proceedings, Ed. V. Kundrat and P. Zavada, 2002, arXiv:hep-ph/0108156.

\bibitem{Meyer:2004jc}
  H.~B.~Meyer and M.~J.~Teper,
  Phys.\ Lett.\ B {\bf 605} (2005) 344.

\bibitem{'tHooft:1973jz}
  G.~'t Hooft,
  Nucl.\ Phys.\ B {\bf 72} (1974)  461.

\bibitem{Malda} J. Maldacena, Adv. Theor. Math. Phys. {\bf 2} (1998) 231.

\bibitem{Wi2} E. Witten, Adv.Theor.Math.Phys. {\bf 2} (1998) 505-532.
  
\bibitem{MASSG} C. Csaki, H. Ooguri, Y. Oz and J. Terning, JHEP {\bf 9901} (1999) 017.

\bibitem{MASSG2} R. de Mello Koch, A. Jevicki, M. Mihailescu , J. P. Nunes,  
Phys.Rev. {D58} (1998) 105009.

\bibitem{MASSG3} A. Hashimoto , Y. Oz , Nucl.Phys. {\bf B548} (1999) 167. 

\bibitem{MASSG4} C. Csaki , Y. Oz , J. Russo , J. Terning , 
Phys.Rev. {D59} (1999) 065012. 

\bibitem{Constable:1999gb}   N.~R.~Constable and R.~C.~Myers,
JHEP {\bf 9910} (1999) 037. 

\bibitem{MASSG5} J. A. Minahan,  JHEP {\bf 9901} (1999) 020.

\bibitem{MASSG6} C. Csaki, J. Terning,  AIP Conf. Proc. 494: 321-328,1999 
(hep-th/9903142).  

\bibitem{MASSG7} R. C. Brower, S. D. Mathur , C. I. Tan , Nucl.Phys. {\bf B587}
(2000) 249. 
 
\bibitem{Caceres:2000qe}
  E.~Caceres and R.~Hernandez,
  Phys.\ Lett.\ B {\bf 504}, 64 (2001)

\bibitem{ACEP}  R. Apreda, D. E. Crooks, N. Evans, M. Petrini, 
JHEP {\bf 0405} (2004) 065.

\bibitem{Amador:2004pz}
  X.~Amador and E.~Caceres,
  JHEP {\bf 0411} (2004) 022.


\bibitem{Evans:2005ip}
  N.~Evans, J.~P.~Shock and T.~Waterson,
Phys.Lett. {\bf B622} (2005) 165.

\bibitem{Caceres:2005yx}
  E.~Caceres and C.~Nunez,
JHEP {\bf 0509} (2005)  027. 

\bibitem{Boschi-Filho:2002vd}
  H.~Boschi-Filho and N.~R.~F.~Braga,
  JHEP {\bf 0305} (2003) 009.

\bibitem{Boschi-Filho:2002ta}
  H.~Boschi-Filho and N.~R.~F.~Braga,
  Eur.\ Phys.\ J.\ C {\bf 32} (2004) 529.

\bibitem{PS} 
J. Polchinski and M. J. Strassler, Phys. Rev. Lett. {\bf  88} (2002) 031601.

\bibitem{Polchinski:2002jw}
  J.~Polchinski and M.~J.~Strassler,
  JHEP {\bf 0305 } (2003) 012.

\bibitem{GI} 
S. B. Giddings, 
Phys.\ Rev.\ {\bf D 67 } (2003) 126001. 

\bibitem{BB3} H. Boschi-Filho and N. R. F. Braga, Phys. Lett. {\bf B560} (2003)
232. 

\bibitem{BT} 
R. C. Brower , C-I Tan, 
Nucl.\ Phys.\ {\bf B 662} (2003) 393.

\bibitem{AN} 
  O.~Andreev,
  Phys.\ Rev.\ D {\bf 67} (2003) 046001.

\bibitem{Brodsky:2003px}
  S.~J.~Brodsky and G.~F.~de Teramond,
  Phys.\ Lett.\ B {\bf 582} (2004)  211.


\bibitem{PandoZayas:2003yb}
  L.~A.~Pando Zayas, J.~Sonnenschein and D.~Vaman,
  Nucl.\ Phys.\ B {\bf 682} (2004) 3.

\bibitem{AN2}
  O.~Andreev,
  Phys.\ Rev.\ D {\bf 70} (2004) 027901.

\bibitem{AN3}
  O.~Andreev,
  Phys.\ Rev.\ D {\bf 71} (2005)  066006.

\bibitem{Andreev:2004sy}
  O.~Andreev and W.~Siegel,
  Phys.\ Rev.\ D {\bf 71} (2005)  086001.


\bibitem{Bigazzi:2004ze}
  F.~Bigazzi, A.~L.~Cotrone, L.~Martucci and L.~A.~Pando Zayas,
  Phys.\ Rev.\ D {\bf 71} (2005) 066002.

\bibitem{Erlich:2005qh}
  J.~Erlich, E.~Katz, D.~T.~Son and M.~A.~Stephanov,
Phys. Rev. Lett. {\bf 95} (2005) 261602.

\bibitem{DaRold:2005zs}
  L.~Da Rold and A.~Pomarol,
 Nucl.\ Phys.\ B {\bf 721} (2005) 79.

\bibitem{GKP} S. S. Gubser , I.R. Klebanov and A.M. Polyakov, 
Phys. Lett. B428 (1998) 105.

\bibitem{Witten:1998qj}
  E.~Witten,
  Adv.\ Theor.\ Math.\ Phys.\  {\bf 2}  (1998) 253.

\bibitem{LAT1} 
C. J. Morningstar and M. Peardon, Phys. Rev. D 56 (1997) 4043; 
Phys. Rev. D60 (1999) 034509;  
``Simulating the scalar glueball on the lattice,''   
AIP Conf.\ Proc.\  {\bf 688} (2004) 220.

\bibitem{LAT2} M.J. Teper, ``Physics from lattice: Glueballs in QCD; 
topology; SU(N) for all N" hep-lat 9711011.

\bibitem{Tang:2000tb}   A.~Tang and J.~W.~Norbury,
  Phys.\ Rev.\ D {\bf 62} (2000) 016006.

\bibitem{Luna:2004gr}
 E.~G.~S.~Luna, M.~J.~Menon and J.~Montanha,
 Nucl.\ Phys.\ A {\bf 745} (2004) 104.

\bibitem{Randall:1999ee}
  L.~Randall and R.~Sundrum,
  Phys.\ Rev.\ Lett.\  {\bf 83} (1999)  3370.

\end{thebibliography}
\end{document}